\documentclass[conference]{IEEEtran}
\usepackage[margin=1.0in]{geometry}
\IEEEoverridecommandlockouts

\usepackage{soul}
\usepackage{xcolor}
\sethlcolor{yellow}
\usepackage{graphicx}
\usepackage{textcomp}

\def\showtrack{0}   
\def\showchange{0}  

\if\showtrack1
    \usepackage{ulem}
    \newcommand\redsout{\bgroup\markoverwith{\textcolor{red}{\rule[0.5ex]{2pt}{0.8pt}}}\ULon}
\else
    \newcommand{\redsout}[1]{}
\fi

\if\showchange1
    
\else
    
\fi

\def\Rtrack{0}   
\def\Rchange{0}  

\if\Rtrack1
    \usepackage{ulem}
    \newcommand\Rredsout{\bgroup\markoverwith{\textcolor{red}{\rule[0.5ex]{2pt}{0.8pt}}}\ULon}
\else
    \newcommand{\Rredsout}[1]{}
\fi

\if\Rchange1
    
\else
    
\fi

\begin{document}
\title{A Secure Dynamic Edge Resource Federation Architecture for Cross-Domain IoT Systems}

\author{
\IEEEauthorblockN{Ronghua Xu${^a}$, Yu Chen${^a}$, Xiaohua Li${^a}$, Erik Blasch${^b}$}
\IEEEauthorblockA{\\${^a}$Dept. of Electrical and Computer Engineering,
Binghamton University, Binghamton, USA \\
${^b}$The U.S. Air Force Research Laboratory, Rome, USA\\
Emails: \{rxu22, ychen, xli\}@binghamton.edu, erik.blasch@us.af.mil}
}

\maketitle

\begin{abstract}
The fast integration of 5G communication, Artificial Intelligence (AI), and Internet-of-Things (IoT) technologies is envisioned to enable Next Generation Networks (NGNs) for diverse smart services and user-defined applications for Smart Cities. However, it is still challenging to build a scalable and efficient infrastructure that satisfies the various performance, security, and management demands by heterogeneous IoT applications across multiple administrative domains. This paper presents a dynamic edge resource federation architecture, which integrates the concept of network slicing (NS) and blockchain to improve scalability, dynamicity, and security for multi-domain IoT applications. A NS-enabled dynamic edge resource federation framework adopts intelligent mechanisms to support efficient multi-domain service coordination that satisfies diverse Quality of Service (QoS) and security requirements. We propose a Hierarchical Integrated Federated Ledger (HIFL), which aims to guarantee decentralized security and privacy-preserving properties in multi-domain resource orchestration and service re-adjustment. As a secure-by-design solution, HIFL is promising to support efficient, trust and secured end-to-end IoT services. A preliminary proof-of-concept prototype has been implemented for comparing intra- and inter-domain performance expectations. 
\end{abstract}


\section{Introduction}
The rapid evolving fifth-generation (5G) communication networks combined with Internet of Things (IoT) and edge-cloud computing technologies brings the concept of \emph{Smart Cities} into practice. A plethora of novel smart applications improve the quality of our lives through enhancing health, safety, and convenience, including intelligent transportation \cite{chen2022anomalous}, smart e-health care \cite{rahmani2015smart}, distributed data markets \cite{xu2021fed}, smart agriculture \cite{xing2021survey}, intelligent edge surveillance \cite{nikouei2019toward}, and more. Meanwhile, smart cities highly rely on an efficient and secure infrastructure, which has to meet various performance, security, and management demands through heterogeneous IoT applications across multiple administrative domains \cite{xu2020blendsps}. 

It becomes more evident that the existing homogeneous network infrastructure is not able to handle scalability and extensibility due to the rapid growth of IoT connections \cite{kassab2020z, srinidhi2019network}. In addition, it is difficult for a traditional centralized service framework to facilitate dynamic and heterogeneous IoT ecosystems that need optimal resource utilization and diverse Quality-of-Service (QoS) requirements \cite{wijethilaka2021survey}. Furthermore, unifying sensitive information among geographically scattered devices that belong to different domains also brings increasingly concerns on security and privacy.  

As one of the key enabling technologies in the context of 5G, Network Slicing (NS) utilizes virtualization and softwarization to divide the physical network into multiple isolated logical networks (i.e, slices) with different network characteristics \cite{wijethilaka2021survey}.
Through dynamic resource allocation among dedicated slices for heterogeneous IoT applications, NS is a promising method to improve scalability and dynamicity in a large scale of IoT network that satisfies  various QoS demands \cite{boubendir2018federation, kafle2018adaptive, taleb2019multi}.
Moreover, the isolation property of NS ensures performance and security guarantees \cite{afolabi2018network}. Thus, NS facilitates mitigating impact of attacks like Distributed Denial-of-Service (DDoS) and protecting sensitive information collected by IoT devices.

Blockchain, a distributed ledger technology (DLT) \cite{nakamoto2019bitcoin}, has demonstrated great potential to revolutionize various aspects of economy and society. Blockchain utilizes a decentralized architecture to securely store and verify data without relying on a centralized trust authority \cite{xu2021decentralized}. The decentralization of blockchain is promising to improve performance and reduce a single point of failure caused by a centralized service architecture. Moreover, leveraging consensus protocols to verify and record transactions on a immutable and transparent public distributed ledger, blockchain guarantees availability, correctness, and provenance for resource sharing (i.e, computing, storage and networking.) among untrusted participants in a multi-domain IoT system.

To improve scalability, dynamicity and security for multi-domain IoT applications, this paper proposes a secure-by-design and dynamic edge resource federation architecture based on NS and blockchain technologies. Integrating NS-enabled dynamic edge resource orchestration with federated ledger fabric \cite{xu2021fed}, the Hierarchical Integrated Federated Ledger (HIFL) solution aims at an organic mutual reinforced networking service infrastructure. All physical edge resources are converted to virtual resources that are managed by domain specific slices. The multi-domain coordination federates these isolated slices that are dynamically designed, deployed and optimized according to service requirements and operating conditions. The federated ledger allows for a decentralized security fabric to enhance security and privacy-preserving properties in intra-slice and inter-slice resource orchestration and service inter-operations.

The remainder of this article is organized as follows. Section \ref{sec:related} introduce the background knowledge and related work of integrating NS and blockchain into IoT systems. In Section \ref{sec:system}, the system architecture of HIFL and main procedures are explained followed with descriptions of the main components and procedures in multi-domain orchestration. Section \ref{sec:prototype} presents the preliminary prototype implementation and evaluation along with discussions on the open questions yet to be addressed. Section \ref{sec:conclusions} concludes this paper. 

\section{Background and Related Work}
\label{sec:related}

\subsection{Network Slicing for IoT }
Network slicing concept in 5G is introduced by NGMN (Next Generation Mobile Network) in \cite{alliance20155g}. To assure service customization, isolation and multi-tenancy support, NS can divide a common physical network infrastructure into multiple logic networks called slices or subnets. Each NS slice is a unification of virtual resources, which run a set of virtual network functions and software defined network (SDN) settings for a specific communication service and business model. NS builds on top of key principles, like isolation, elasticity, automation and customisation, which are promising to improve scalability and dynamicity of heterogeneous IoT systems \cite{afolabi2018network}.

To support the intelligent allocation and dynamic adjustment of virtual resources in NS networks, adaptive virtual network slices for diverse IoT services is proposed to achieve vertical, horizontal and internetwork scaling purposes \cite{kafle2018adaptive}. Given real-time resource utilization and performance requirements, adaptive resource adjustment algorithms can produce the output of new optimal values to re-allocate resources. 
NS may combine resources from different administrative domains. Thus, a multi-domain network slicing management and orchestration architecture is designed to enable efficient and dynamic federated resources allocation across domains \cite{taleb2019multi}. Each domain uses sub-domain controllers to orchestrate and manage resources and Network Slice Subnet Instances (NSSI), while an global end-to-end slice coordinator unifies the management of federated domains. 
By using a brokering layer that relies on a graph-based resource database, a brokering architecture for network slicing is proposed to federate IT and edge resources owned by multiple third-party actors \cite{boubendir2018federation}. The proposed architecture \cite{boubendir2018federation} is implemented for the federation of a stadium infrastructure resources, and an author evaluates the time for resource federation, slice provisioning and slice activation.

\subsection{IoT-Blockchain Considerations}
\textit{Blockchain} initially was implemented as an enabling technology of Bitcoin \cite{nakamoto2019bitcoin}, which aims to provide a cryptocurrency to record and verify commercial transactions among trustless entities in a decentralized manner. In a blockchain network, a large amount of miners or validators execute a consensus protocol under a Peer-to-Peer (P2P) network to ensure integrity, consistence and total order of data on the distributed ledger. Thanks to the decentralized network architecture and cryptographic security mechanisms, all trust-less participants in a blockchain system cooperatively maintain a security and trust framework instead of relying on a centralized third party trust authority. Emerging from the intelligent property, a \textit{smart contract} (SC) encapsulates self-executing procedures recorded on the distributed ledger, Thus, a SC introduces programmability into blockchain to support various customized transaction logic rather than simple cash transactions \cite{xu2020blendsps}.

Combining blockchain and a smart contract enables a secured and trust-free framework to facilitate data sharing and the federation of resources from third party actors. However, resources intensive consensus algorithms, like Proof-of-Work (PoW) and its variants, are not affordable for IoT devices that are strictly constrained by computation and storage capacity. Using a Practical Byzantine Fault Tolerance (PBFT) \cite{castro1999practical} protocol can achieve high throughput, lower latency and limited computation overhead; however, it cannot scale up to a large consensus network. Moreover, IoT devices are managed by different administrative domains with diverse performance and security requirements. Therefore, a monolithic blockchain network cannot perfectly ensure decentralization, scalability and security for dynamic and heterogeneous IoT systems across multiple domains. 

\subsection{Lightweight Distributed Ledgers for Edge Networks}

The inherent security guarantees of blockchain provide the foundations of a serverless record-keeping without the need for centralize trusted third-party authorities \cite{ali2018applications}. Transparency, immutability and auditability ensure resilience, correctness, and provenance for all data sharing among untrusted participants. Many efforts try to leverage blockchain to support security features required in IoT systems. IoTChain \cite{bao2018iotchain} proposes a three-tier blockchain-based IoT architecture, which allows regional nodes to perform any lightweight consensus, like Proof-of-Stake (PoS) and PBFT. IoTChain only provides simulation results on communication cost of transactions; however key metrics in the consensus layer, like computation, storage and throughput, are not considered. FogBus \cite{tuli2019fogbus} proposes a lightweight framework for integrating blockchain into fog-cloud infrastructure, which aims to ensure data integrity as transferring confidential data over IoT-based systems. In FogBus, master nodes deployed at the fog layer are allowed to perform PoW mining, while IoT devices send transactions to master nodes as trust intermediates to interact with blockchain. However, using PoW as the backbone consensus protocol still results in high energy consumption and low throughout.

HybridIoT \cite{sagirlar2018hybrid} proposes a hybrid blockchain-IoT architecture to improve scalability and interoperability among sub-blockchains. In HybridIoT, a BFT inter-connector framework works as a global consortium-blockchain to link multiple PoW sub-blockchains. However, using PoW consensus in sub-blockchain networks still brings computation and storage overhead on IoT devices if they are deployed as full nodes.
IoTA \cite{iota} aims to enable a cryptocurrency designed for the IoT industry, and it leverages a directed acyclic graph (DAG), called tangle \cite{popov2016tangle}, to record transactions rather than chained structure of the ledger. IoTA provides a secure data communication protocol and zero fee micro-transaction for IoT/machine-to-machine (M2M), and it demonstrates high throughput and good scalability. However, existing IoTA networks still rely on hard-coded coordinators, which employ PoW to finalize path of recorded transactions in DAG.

Unlike the above mentioned IoT-Blockchain solutions, which either adopt computation intensive PoW as their backbone consensus mechanism or rely on a intermediate fog layer to execute consensus protocol, HIFL aims at a partially decentralized, lightweight hierarchical blockchain network fabric that is customized for a dynamic network slice in a highly heterogeneous edge computing environment. In addition, HIFL will form an organic network fabric of the network slices, not merely taking advantage of blockchain as a time-stamped series of data records, but can also serve as network slice brokers \cite{nour2019blockchain, valtanen2018creating, zanzi2020nsbchain}.
\section{HIFL: Rationale and Design}
\label{sec:system}

\subsection{System Architecture Design}
\begin{figure*} [t]
\begin{center}
\begin{tabular}{c}
\includegraphics[height=8.5 cm]{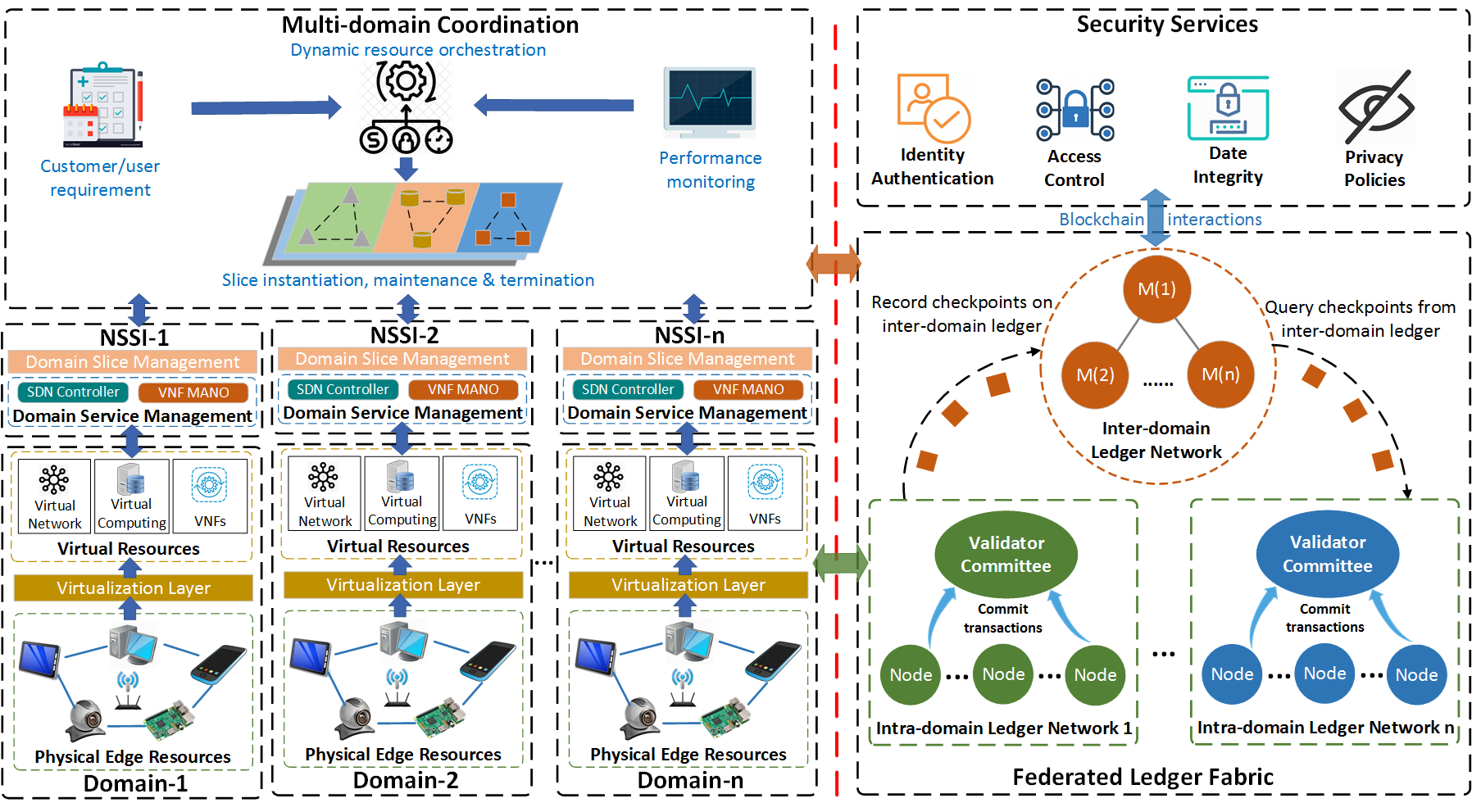}
\end{tabular}
\end{center}
\vspace{-0.1in}
\caption[example] { \label{fig:architecture} Illustration of system architecture consisting of multi-domain service coordination and federated ledger fabric.}
\end{figure*}

Aiming at a self-adaptive, secure-by-design and partial decentralized networking service architecture, the HIFL solution takes advantage of NS and blockchain technologies to enable efficient and scalable edge resources federation and orchestration under heterogeneous multi-domain IoT environments.
The proposed HIFL architecture is depicted in Figure \ref{fig:architecture} with two distinct sub-frameworks.

(1) \emph{Multi-domain Coordination}: adopts a dynamic edge resource federation paradigm with NS as an enabling technology. The virtualization layer abstracts all physical edge resources to virtual resources according to functionalities, like computing, storage and network connectivity.
Each domain specific slice manages virtual resources within a domain and provides interfaces to upper level multi-domain coordination. A software-defined network (SDN) controller provides network connectivity and service chaining among allocated Virtual Networ Functions (VNF). VNF Management and Orchestration (MANO) manages the VNFs along with required virtual computing and storage resources.
Dynamic resource orchestration depends on customer/user requirements and system performance monitoring. By federating virtual resources managed by different domain specific slices, multi-domain orchestration relies on intelligent algorithms to achieve fast resource deployment and efficient services re-adjustments with diverse QoS and security requirements. 

(2) \emph{Federated Ledger}: provides a decentralized security fabric to guarantee security and privacy-preserving in data and resource sharing across different domains. In a specific domain, a random elected committee executes an efficient BFT-based consensus protocol to verify and record data on a private intra-domain ledger. Because each domain is a permissioned network, only authorized users are allowed to access data on an intra-domain ledger. Moreover, running BFT consensus protocol by a small scale validator committee can achieve low latency and high throughput of transactions. Such an intra-domain ledger blockchain network supports partial decentralization at the network of edge with performance and privacy-preserving guarantees.
At the multi-domain level, a public inter-domain ledger network federates fragmented intra-domain ledger networks, and it uses a scalable PoW consensus to secure cross-domain operations. For multi-domain operations, the public inter-domain ledger only records checkpoints data that indirectly refer to raw data on private intra-domain ledgers. Therefore, the NIFL federated ledger structure is promising to ensure scalability and security without sacrificing performance and privacy requirements of individual domains \cite{xu2021fed}.

\subsection{Dynamic Resource Orchestration}

\begin{figure}[t]
    \begin{center}
    \includegraphics[width=0.47\textwidth]{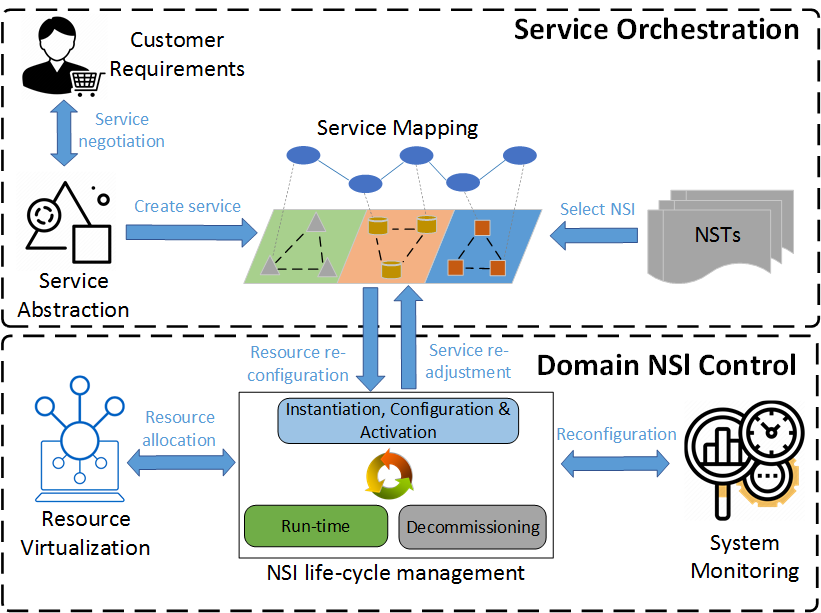}
    \vspace{-0.1in}
    \caption { \label{fig:workflow} The dynamic resource orchestration mechanism.} 
    \end{center}
\end{figure}

The dynamic resource orchestration uses federated NS Instances (NSIs) to support end-to-end connectivity for multi-domain applications. Given continuously analyzing service and monitoring performance, NSIs carry out resource allocation and service re-adjustment to ensure desired requirements. An overview of multi-domain service orchestration and domain NSI control is shown in Figure \ref{fig:workflow}. The dynamic resource orchestration happens if 1) customers/users launch service requests; or 2) running services with allocated resources cannot fulfill desired performance.

1) After receiving customer requirements, Service Orchestration (SO) firstly negotiates with verticals and service providers on admission control and charging. Then, SO decomposes and abstracts services toward different administrative domains. Given current unified resources of the system, service mapping process selects appropriate NS templates (NSTs) to create a service graph that is sent to domain NSI control for resource configuration. The domain NSI control allocates virtual resources for new NS instantiation. After configuration and activation, an NSI becomes run-time that supports functionalities of service and reports performance.

2) When system monitoring detects performance degradation due to insufficient resource or service configurations, domain NSI's control will send service re-adjustment requests to service orchestration. The service mapping process may modify service specific parameters, and/or allocate more topology, links and computing resources. Then it sends resource re-configuration to domain NSI control, which is responsible for instantiating, modifying or decommissioning NS to meet on-demand requirements in service providing time.

\subsection{Blockchain based Security Mechanism}

\begin{figure*} [t]
\begin{center}
\begin{tabular}{c}
\includegraphics[height=4.2 cm]{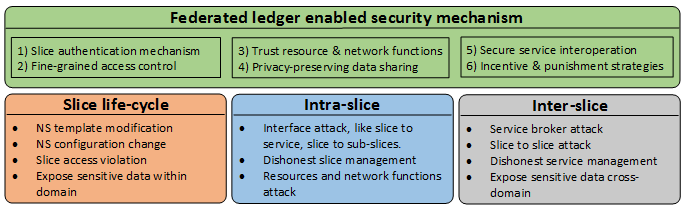}
\end{tabular}
\end{center}
\caption[example] { \label{fig:security} Blockchain based security mechanism for dynamic edge resource federation.}
\end{figure*}

The identified threads and probable points of attack for network slicing can be categorized as: life-cycle security, intra-slice security and inter-slice security \cite{olimid20205g}.
Figure \ref{fig:security} summarizes the representative threats and explains how blockchain can be used to enhance security proprieties of dynamic edge resource federation framework.

\emph{Slice life-cycle Attack}: An adversary can modify NS templates in preparation phase or change configuration when new slices are instantiating given on-demand service requests. The access violation may happen in slice run-time phases such that unauthorized entities can conduct performance and DoS attacks by changes in configuration or even deactivation of slices. Moreover, improper decommissioning handles may expose sensitive data of resource providers within a domain.

For each domain assuming that the domain administrator is a trust oracle, a dedicated intra-domain ledger network relies on permissioned management to provide flexible authentication primitives, like identity verification \cite{xu2019exploration} and access control \cite{xu2018blendcac}. It is promising to prevent against unauthorized access to data and resource as well as performing malicious operations under a dynamic network condition. Moreover, slice template and instance information can be recorded on an immutable and auditable intra-domain ledger storage. Thus, a decentralized data integrity scheme \cite{nikouei2018real} supports authenticity and integrity verification for slices, to prevent fake or modified slice instances when new resources join or leave the system. 

\emph{Intra-slice Attack}: Slices within a domain also expose vulnerabilities on interfaces between services and sub-slices. Moreover, a slice manager collects resources and network functions shared by different providers. Thus, it needs audit participants' behaviors to identify dishonest activities. The intra-domain ledger provides a trust free platform that all authorized participants can verify data on the distributed ledger without relying on any third-party agency. By properly designing authentication and access control strategies, the intra-domain ledger can guarantee privacy-preserving data and resource sharing among untrusted providers.

\emph{Inter-slice Attack}: Considering inter-slice scenarios, an adversary can compromise service brokers to interrupt multi-domain coordination, or use authorized slices to gain access to other unauthorized slices. Moreover, it may leak sensitive data during inter-slice operations. The HIFL federated ledger leverages a cryptographic secure inter-ledger transaction (tx) protocol to guarantee auditability and provenance of multi-domain tasks without exposing sensitive data on intra-domain ledgers. Moreover, incentives and punishment strategies by blockchain motivate more honest participants to join the system and gain benefits in data and resources sharing. 

\section{Proof-of-Concept Prototype Evaluation}
\label{sec:prototype}

\begin{table}[t]
\vspace{-0.10in}
\caption{Configuration of Experimental Devices.} 
\vspace{-0.18in}
\label{tab:testbed}
\begin{center}       
\begin{tabular}{|l|p{2.5cm}|p{2.8cm}|} 
\hline
\rule[-1ex]{0pt}{3.5ex} \textbf{Device} & Dell Optiplex-7010 & Raspberry Pi 4 (B) \\
\hline
\rule[-1ex]{0pt}{3.5ex} \textbf{CPU} & Intel Core TM i5-3470 (4 cores), 3.2GHz & Broadcom ARM Cortex A72 (ARMv8), 1.5GHz \\
\hline
\rule[-1ex]{0pt}{3.5ex} \textbf{Memory} & 8GB DDR3 & 4GB SDRAM \\
\hline
\rule[-1ex]{0pt}{3.5ex} \textbf{Storage} & 350G HHD & 64GB (microSD card) \\
\hline
\rule[-1ex]{0pt}{3.5ex} \textbf{OS} & Ubuntu 16.04 & Raspbian GNU/Linux (Jessie) \\
\hline
\end{tabular}
\end{center}
\end{table}

To study the feasibility of proposed HIFL solution, we implemented a conceptual prototype of resource federation that simulates a video surveillance system including two administrative domains. Table \ref{tab:testbed} shows configuration of devices for prototype setup.
Each domain consists of 10 Raspberry Pi-4 (RPi) devices that provide edge resources and a Dell Optiplex-7010 desktop as a domain manager.
We use Tendermint core \cite{tendercore} to build a intra-ledger network for each domain, and
all RPis within a domain also act as validators to execute an efficient BFT protocol and maintain its intra-domain ledger.
A private Ethereum network is used to simulate a inter-domain ledger network, where 4 miners are deployed on separate desktops and each has Intel(R) Core(TM) 2 Duo CPU E8400 @ 3 GHz and 4 GB of RAM.
All desktops and RPis are connected through a local area network (LAN).

\subsection{Numerical Results}
Table \ref{tab:comparison} provides a set of comparative numeral results by querying or recording data on intra-domain and inter-domain networks separately. 100 test runs have been conducted for each test scenario and we use the average value to show performance. Because querying data from an inter-domain ledger uses smart contract as enable technology, such that it needs more processing time than intra-domain ledger does. In addition, the HIFL intra-domain ledger relies on an efficient BFT consensus protocol, which achieves lower latency and higher transaction throughput as recording data on distributed ledger. Thus, HIFL is able to satisfy time sensitive and high throughput requirements in specific domain networks. Moreover, the inter-domain ledger is mainly to guarantee scalability, auditability and global security for cross-domain operations. Therefore, 4.5 s transaction latency and 127 tx/s transaction rate is acceptable for majority of inter-domain scenarios.

To evaluate resource consumption as validators and miners run on host machines, we use ``top'' command to monitor CPU and memory usages by consensus processes on desktop and RPi.
The Ethereum miner uses a computation intensive PoW mining algorithm such that needs full capacity of a  CPU core and consumes about 1.2GB memory. As a result, miners can only be deployed on the powerful platforms, like edge or fog servers.
Due to lightweight BFT consensus algorithm that achieves efficiency in CPU and memory usage, it's affordable for IoT devices to work as validators in intra-ledger networks. 
Ethereum network requires gas fees that are used to reward miners who commit transactions on the inter-ledger. It introduces extra financial cost on inter-ledger transactions, which is \$1.23/$tx$ according to Ether price of the public Ethereum market at Jan 22, 2021.
However, intra-ledger transactions do not require transaction fees, and each domain can design their own rewarding strategies for participants.

\begin{table}[t]
\caption{Comparative evaluation of committing transactions on federated ledger networks.} 
\label{tab:comparison}
\begin{center}       
\begin{tabular}{|l|c|c|} 
\hline
\rule[-1ex]{0pt}{3.5ex}  & Intra-domain & Inter-domain \\
\hline
\rule[-1ex]{0pt}{3.5ex} \textbf{query data (ms)} &   18  &  110 \\
\hline
\rule[-1ex]{0pt}{3.5ex} \textbf{record data (s)} &   1.6  &  4.5 \\
\hline
\rule[-1ex]{0pt}{3.5ex} \textbf{$tx$ throughput (tx/s)} &  625  &  127 \\
\hline
\rule[-1ex]{0pt}{3.5ex} \textbf{CPU usage (\%)} &  100  &  32 \\
\hline
\rule[-1ex]{0pt}{3.5ex} \textbf{Memory usage (MB)} &  1,200  &  80 \\
\hline
\rule[-1ex]{0pt}{3.5ex} \textbf{Gas/$tx$ (Ether) } &  0.001  &  $\times$ \\
\hline
\end{tabular}
\end{center}
\end{table}

\subsection{Discussions}
This paper focuses on the design rationale and principles of the HIFL solution. In terms of completeness, it is an introduction of a work-in-progress. There are a lot of open questions yet to be answered and limitations of the current proof-of-concept prototype. 

Basically, HIFL takes advantages of properties of NS and blockchain to achieve multi-domain dynamic edge resources orchestrations and service re-adjustments on a physically distributed infrastructure. Actually HIFL still relies on a partially decentralized network framework owing to a logically centralized administrative management for domain-specific NS and VNF. 

In addition, HIFL is a promising basic abstract architecture to address issues in distributed, dynamic and heterogeneous cross-domain IoT systems. Because of the high diversity in the smart cities application domain and the design of NS itself is highly application dependent, implementations based on certain specific real-life applications are necessary to gain deeper insights and demonstrate practical benefits of using HIFL.
On top of a preliminary proof-of-concept prototype that is built leveraging our earlier work on intelligent public safety surveillance, we have obtained some numerical results that evaluate general performance in terms of network latency and processing overhead. Due to limited efforts, however, we have not validated security and privacy properties given possible attack scenarios. We leave aforementioned issues to future work.

\section{Conclusions}
\label{sec:conclusions}

This paper proposes HIFL - a partially decentralized and secure-by-design dynamic edge resource federation architecture atop of multi-domain IoT systems. Experimental results based on a proof-of-concept prototype demonstrate the feasibility of applying the HIFL solution in smart cities scenarios, like smart video surveillance systems. While the HIFL experimental results are encouraging, what we have presented in this paper is merely an initial conceptual design and some preliminary study results. There are many open questions are yet to be solved before the HIFL architecture becomes applicable in real-world smart cities applications. Our current ongoing efforts mainly focus on the design of an efficient dynamic edge resource orchestration mechanism and an approach for the evaluation on the performance and security features in a large scale IoVT network.

\section*{Acknowledgement}

This work is supported by the U.S. National Science Foundation (NSF) via grant CNS-2141468 and the U.S. Air Force Office of Scientific Research (AFOSR) Dynamic Data and Information Processing Program (DDIP) via grant FA9550-21-1-0229. The views and conclusions contained herein are those of the authors and should not be interpreted as necessarily representing the official policies or endorsements, either expressed or implied, of the U. S. Air Force. 

\bibliographystyle{IEEEtranS} 
\bibliography{References} 

\end{document}